\documentclass[aps,prc,preprint,showpacs,floatfix]{revtex4-1}  
\usepackage{graphicx}
\usepackage{amsmath}
\usepackage{physics}
\newcommand {\nc} {\newcommand}
\nc {\beq} {\begin{eqnarray}} \nc {\eol} {\nonumber \\} \nc {\eeq}
{\end{eqnarray}} \nc {\eeqn} [1] {\label{#1} \end{eqnarray}} \nc
{\eoln} [1] {\label{#1} \\} \nc {\ve} [1] {\mbox{\boldmath $#1$}}
\nc {\rref} [1] {(\ref{#1})} \nc {\Eq} [1] {Eq.~(\ref{#1})} \nc
{\re} [1] {Ref.~\cite{#1}} \nc {\dem} {\mbox{$\frac{1}{2}$}} \nc
{\arrow} [2] {\mbox{$\mathop{\rightarrow}\limits_{#1 \rightarrow
#2}$}}

\begin{document}
\title{New method for the solution of the two-body Dirac equation
 for the positronium bound states}

\author {E.M. Tursunov}
\email{tursune@inp.uz} \affiliation {Institute of Nuclear Physics,
Academy of Sciences, 100214, Ulugbek, Tashkent, Uzbekistan}
\affiliation {National University of Uzbekistan, 100174  Tashkent,
Uzbekistan}
\author {Sh.G. Norbutaev}
\email{norbutaev@inp.uz} \affiliation {Institute of Nuclear Physics,
Academy of Sciences, 100214, Ulugbek, Tashkent, Uzbekistan}
\author {B.A. Fayzullaev}
\email{bfayzullaev@nuu.uz} \affiliation {National University of
Uzbekistan, 100174 Tashkent, Uzbekistan}

\begin{abstract}
A new theoretical method is developed for the solution of the
two-body bound-state Dirac equation for positronium. Only Coulomb
potential was included in the Dirac Hamiltonian. It is shown that
the two-body Dirac Hamiltonian can be written in the Hermitian
matrix form of the $4\times 4$ size and diagonalized in the
momentum-state representation. Numerical results for the energy
spectrum of the para- and ortho-positronium ground states performed
within the variational method using the harmonic oscillator basis
functions are in good agreement with a high-precision finite-element
method of T.C. Scott et al. After the Fourier transformation into
the coordinate-state representation the bound state wave functions
of the para-Ps and ortho-Ps do not contain any singularity at the
origin in contrast to above mentioned method. The weights of the
large-small and small-large components of the ground state wave
functions are estimated to be of order 10$^{-6}$, while the weight
of the small-small component is of order 10$^{-12}$.
\end{abstract}


 \maketitle

\section{Introduction}
\par A new generation of the high sensitivity and multi-photon total-body positron emission tomography
systems opens perspectives for clinical applications of positronium
(Ps) in medicine \cite{bas2023}. On the other hand, Ps is the most
important object for the precision tests of the violation of the
discrete symmetries such as {\it C} (charge conjugation), {\it P}
(spatial parity), {\it T} (time reversal) and their combinations
\cite{PhysRep2022,vet2002,rub2004,yam2010,bar2020,hau2023,NatCom2024}.
The spin-singlet bound states of the electron $e^-$ and its
antiparticle, the positron $e^+$ present para-positronium ($p$-Ps),
at the same time the spin-triplet bound states present the
ortho-positronium ({\it o}-Ps).

The most important spectroscopic properties of the positronium, such
as energy-level intervals are reproduced with a high precision at
order $\alpha^6$ within the non-relativistic QED
\cite{PR13,PR25,PR125,PR126,PR242} based on the solutions of the
Schr\"{o}dinger bound-state equation. The explicit relativistic
invariance built into QED is essential for particle physics but
leads to unnecessary complications when QED is applied to Coulombic
bound states, which deals with the energy scales small compared to
$m_ec^2$ \cite{PhysRep2022}.

The two-body Dirac equation (TBDE) for positronium was studied in
Ref. \cite{scott1992} using only the Coulomb interaction potential.
In Ref. \cite{fer2023} the sixteen-component, no-pair
Dirac-Coulomb-Breit equation, derived from the Bethe-Salpeter
equation, is solved in a variational procedure using Gaussian-type
basis functions for positronium, muonium, hydrogen atom, and muonic
hydrogen. Another type of relativistic Schr\"{o}dinger equation was
obtained and solved based on the Pauli reduction of the TBDE
\cite{cra2006,cra2009}.

Recent studies of properties of anomalous bound states within the
TBDE formalism \cite{pat19} have demonstrated that there are still
many open questions for the structure of positronium. It was shown
that the radial components of the Positronium bound-state
wave-function \cite{scott1992} contain a singularity near the origin
and differ from their Pauli approximations. This difference is due
to mixing of the ground-state wave function with the anomalous
bound-state wave functions near the origin. The solutions of the
Bethe-Salpeter equation do not show any mixing with anomalous states
and do not have a singularity near the origin \cite{pat23}.

The question is whether a bound state solution of the two-body Dirac
equation with a pure Coulomb interaction potential can be free of
mixing with an anomalous state. The aim of present work is to study
this question within an alternative method. We solve the TBDE for
the positronium bound states within a new theoretical method. The
method is based on the Hermitian Dirac Hamiltonian in the form of
$4\times 4$ matrix, each element of which represents a
four-component operator. The solution of the TBDE for the
positronium state with total momentum $J$ and its projection $M$
with the spin-orbital couplings
$\vec{\mathbf{J}}=\vec{\mathbf{l}}+\vec{\mathbf{S}}$ and
$\vec{\mathbf{S}}=\vec{\mathbf{S}}_1+ \vec{\mathbf{S}}_2$,
$S_1=S_2=1/2$, can be expressed in the form of a column of four
components. Each of the large-large (LL), large-small (LS),
small-large (SL) and small-small (SS) components of the eigen
function represents a four-component spinor. The radial parts of
these components are expressed as linear combination of the basic
functions. This form of the Hermitian Dirac Hamiltonian contains a
term $\vec{\boldsymbol{\sigma}} \hat{\vec{\mathbf {p}}}$,
responsible for the coupling of the main large-large component with
the large-small and small-large components. Numerical calculations
have been performed within the variational method on the harmonic
oscillator basic functions. The matrix elements of the most
important operators including $\vec{\boldsymbol{\sigma}}
\hat{\vec{\mathbf {p}}}$, are calculated in momentum state
representation and are all real numbers in contrast to the
coordinate space representation. As a results we come to the system
of algebraic equations which is solved numerically
\cite{gutsche,gutschedis,tur09,tur14}.

In Section II we give the main formalism of the method. Section III
deals with numerical results and the conclusions are presented in
the last Section.
\section{Solution of the two-body bound state Dirac equation}
\subsection{Two-body Dirac equation for  positronium bound states}
    \qquad The Dirac Hamiltonian for the two-body bound system
    interacting via Coulomb potential ($\hbar=c=1$)   is written in the form \cite{scott1992}
    \begin{equation}\label{hamil}
    \hat{\mathbf{H}}=\vec{\boldsymbol{\alpha}}_1\times\mathbf{I}^{(2)}_{4}\cdot\vec{\mathbf{p}}_1+
    \mathbf{I}^{(1)}_4\times\vec{\boldsymbol{\alpha}}_2\cdot\vec{\mathbf{p}}_2+\beta_1\times\mathbf{I}^{(2)}_4m_1+
    \mathbf{I}^{(1)}_4\times\beta_2m_2+\mathbf{I}_{16}{V}(r_{12}),
    \end{equation}
    where the Dirac matrices
\begin{equation}
    \vec{\boldsymbol{\alpha}}=\beta\vec{\boldsymbol{\gamma}}=
    \begin{pmatrix}
        0&\vec{\boldsymbol{\sigma}}\\\vec{\boldsymbol{\sigma}}&0
    \end{pmatrix},
    \qquad\beta=\gamma^0=
    \begin{pmatrix}
        \mathbf{I}_2&0\\0&-\mathbf{I}_2
    \end{pmatrix},\end{equation}
and  $\mathbf{I}_{n}$ is the unit $n\times n$ matrix operator,
${V}(r_{12})$ is the Coulomb potential. Index 1 (2) of the operator
means that this operator acts on the wave function of the first
(second) particle. In order to write the Hamiltonian in the relative
coordinates we introduce the radius-vectors of the center of mass
$\vec{\mathbf{R}}$  and the relative-motion $\vec{\mathbf{r}}$  of
the positron-electron two-body system:
\begin{equation*}\begin{array}{l}
        \vec{\mathbf{R}}=\dfrac{1}{2}(\vec{\mathbf{r}}_1+\vec{\mathbf{r}}_2), \quad \vec{\mathbf{r}}=\vec{\mathbf{r}}_1-\vec{\mathbf{r}}_2.
\end{array}\end{equation*}
With the assumption $\vec{\mathbf{P}}_{\mathbf{R}}=0$, one can write
the total Dirac Hamiltonian of the relative motion in terms of the
relative momentum $\vec{\mathbf{p}}$ and coordinate
$\vec{\mathbf{r}}$:
\begin{equation}\begin{array}{l}\label{hamilt}
        \hat{\mathbf{H}}=\left\{\vec{\boldsymbol{\alpha}}_1\times\mathbf{I}^{(2)}_4-\mathbf{I}^{(1)}_4\times\vec{\boldsymbol{\alpha}}_2\right\}\cdot\vec{\mathbf{p}}+
        \beta_1\times\mathbf{I}^{(2)}_4m_1+\mathbf{I}^{(1)}_4\times\beta_2m_2+\mathbf{I}_{16}{V}(r).
\end{array}\end{equation}
Using the standard representation of the Dirac gamma matrices, the
kinetic part of the Hamiltonian can be written in the form of
symmetric matrix:
\begin{equation}
    \left\{\vec{\boldsymbol{\alpha}}_1\times \mathbf{I}^{(2)}_{4}-\mathbf{I}^{(1)}_{4}\times\vec{\boldsymbol{\alpha}}_2\right\}\cdot\vec{\mathbf{p}}=
    \begin{pmatrix}
        0&-\mathbf{I}^{(1)}_{2}\times\vec{\boldsymbol{\sigma}}_2\cdot\vec{\mathbf{p}}&\vec{\boldsymbol{\sigma}}_1\times \mathbf{I}^{(2)}_{2}\cdot\vec{\mathbf{p}}&0\\\\-\mathbf{I}^{(1)}_{2}\times\vec{\boldsymbol{\sigma}}_2\cdot\vec{\mathbf{p}}&0&0&\vec{\boldsymbol{\sigma}}_1\times \mathbf{I}^{(2)}_{2}\cdot\vec{\mathbf{p}}\\\\\vec{\boldsymbol{\sigma}}_1\times \mathbf{I}^{(2)}_{2}\cdot\vec{\mathbf{p}}&0&0&-\mathbf{I}^{(1)}_{2}\times\vec{\boldsymbol{\sigma}}_2\cdot\vec{\mathbf{p}}\\\\0&\vec{\boldsymbol{\sigma}}_1\times \mathbf{I}^{(2)}_{2}\cdot\vec{\mathbf{p}}&-\mathbf{I}^{(1)}_{2}\times\vec{\boldsymbol{\sigma}}_2\cdot\vec{\mathbf{p}}&0
    \end{pmatrix}.
\end{equation}
    Then the total Hermitian Dirac Hamiltonian can be written in the form of $4\times4$ matrix, each element of which represents the  $4\times4$ matrix operator acting on the
    four-component spinor:
\begin{equation}
    \hat{\mathbf{H}}=
    \begin{pmatrix}
        \boldsymbol{\delta}_1+\mathbf{I}_4{V}(r_{12})&-\mathbf{I}^{(1)}_2\times\vec{\boldsymbol{\sigma}}_2\cdot\vec{\mathbf{p}}&\vec{\boldsymbol{\sigma}}_1\times\mathbf{I}^{(2)}_2\cdot\vec{\mathbf{p}}&0\\\\-\mathbf{I}^{(1)}_2\times\vec{\boldsymbol{\sigma}}_2\cdot\vec{\mathbf{p}}&\boldsymbol{\delta}_2+\mathbf{I}_4{V}(r_{12})&0&\vec{\boldsymbol{\sigma}}_1\times\mathbf{I}^{(2)}_2\cdot\vec{\mathbf{p}}\\\\\vec{\boldsymbol{\sigma}}_1\times\mathbf{I}^{(2)}_2\cdot\vec{\mathbf{p}}&0&-\boldsymbol{\delta}_2+\mathbf{I}_4{V}(r_{12})&-\mathbf{I}^{(1)}_2\times\vec{\boldsymbol{\sigma}}_2\cdot\vec{\mathbf{p}}\\\\0&\vec{\boldsymbol{\sigma}}_1\times\mathbf{I}^{(2)}_2\cdot\vec{\mathbf{p}}&-\mathbf{I}^{(1)}_2\times\vec{\boldsymbol{\sigma}}_2\cdot\vec{\mathbf{p}}&-\boldsymbol{\delta}_1+\mathbf{I}_4{V}(r_{12})
    \end{pmatrix},
    \label{Hamiltonian}
\end{equation}
where the diagonal elements read
$\boldsymbol{\delta}_1=\mathbf{I}^{(1)}_2\times\mathbf{I}^{(2)}_2(m_1+m_2)$,
\quad
$\boldsymbol{\delta}_2=\mathbf{I}^{(1)}_2\times\mathbf{I}^{(2)}_2(m_1-m_2)$,
\quad ${V}(r_{12})=-\dfrac{\alpha}{r}$. 
The identity operators  $\mathbf{I}_2^{(1)} $, $\mathbf{I}_2^{(2)}$
of the $2\times2$ size act only on the spin function of the first
and second particles, respectively. The above form of the Hermitian
Dirac Hamiltonian matrix differs from the non-Hermitian matrix
Hamiltonian of Ref. \cite{scott1992}. However, they coincide with
each other in the $16\times16$ matrix form.

It can be proven that the operators
$-\mathbf{I}_{2}^{(1)}\times\vec{\boldsymbol{\sigma}}_2\cdot
\hat{\vec{\mathbf{p}}}$ and
$\vec{\boldsymbol{\sigma}}_1\times\mathbf{I}_{2}^{(2)}\cdot\hat{\vec{\mathbf{p}}}$
of the Dirac Hamiltonian Eq.(\ref{Hamiltonian}), where
$\hat{\vec{\mathbf{p}}}=\vec{\mathbf{p}}/ p$ is the unit momentum
vector, while acting on the spherical tensor
$\mathcal{Y}^{{JM}}_{\ell{S}}(\vec{\xi}_1,\vec{\xi}_2;\hat{\vec{\mathbf{p}}})$,
can change the spin $S$ and the orbital momentum $\ell$ of the
system by one:
\begin{equation}\begin{array}{l}
        (-\mathbf{I}^{(1)}_{2}\times\vec{\boldsymbol{\sigma}}_2\cdot\hat{\vec{\mathbf{p}}})\mathcal{Y}^{{JM}}_{({S}_1,{S}_2){S}\ell}(\vec{\xi}_1,\vec{\xi}_2;
        \hat{\vec{\mathbf{p}}})=(-1)^{\ell+\ell^{\pm}}\sum_{h,{S}'}[h](\left[{S}\right]\left[{S}'\right])^{\frac{1}{2}}\left\{\begin{array}{ccc}
         S_1&S_2&{S}\\\ell&{J}&h\end{array}\right\}\\\\\cdot\left\{\begin{array}{ccc}
          S_1&S_2&{S}'\\\ell^{\pm}&{J}&h\end{array}\right\}\mathcal{Y}^{{JM}}_{({S}_1,{S}_2){S}'\ell^{\pm}}(\vec{\xi}_1,\vec{\xi}_2;
          \hat{\vec{\mathbf{p}}}),
\end{array}\label{important}\end{equation}
\begin{equation}\begin{array}{l}
        (\vec{\boldsymbol{\sigma}}_1\times\mathbf{I}^{(2)}_{2}\cdot\hat{\vec{\mathbf{p}}})\mathcal{Y}^{{JM}}_{({S}_1,{S}_2){S}\ell}(\vec{\xi}_1,\vec{\xi}_2;
        \hat{\vec{\mathbf{p}}})=(-1)^{\ell+\ell^{\pm}+1}\sum_{h,{S}'}(-1)^{{S}+{S}'}[h]\left(\left[{S}\right][{S}']\right)^{\frac{1}{2}}\left\{\begin{array}{ccc}
           S_1&S_2&{S}\\\ell&{J}&h\end{array}\right\}\\\\\cdot\left\{\begin{array}{ccc}
            S_1&S_2&{S}'\\\ell^{\pm}&{J}&h\end{array}\right\}\mathcal{Y}^{{JM}}_{({S}_1,{S}_2){S}'\ell^{\pm}}
            (\vec{\xi}_1,\vec{\xi}_2;\hat{\vec{\mathbf{p}}}),\\
\end{array}\label{important1} \end{equation}
where $\ell^{\pm}=\ell {\pm}1$, $S'=0$ and/or 1, and $\vec
{\boldsymbol{\ell}^{\pm}}+\vec{\mathbf{S}}'=\vec{\mathbf{ J}}$. The
last equations Eq.(\ref{important}) and Eq.(\ref{important1})
contain the most important result for the structure of positronium
within the two-body Dirac equation formalism.
 \subsection{Variational method on a harmonic-oscillator basis}
The solution of the two-body Dirac equation for the positronium
state with total momentum $J$ and its projection $M$, where
$\vec{\mathbf{J}}=\vec{\boldsymbol{\ell}}+\vec{\mathbf{S}}$\quad
and\quad $\vec{\mathbf{S}}=\vec{\mathbf{S}}_1+\vec{\mathbf{S}}_2$,
\\
\begin{equation}
    \left(\hat{\mathbf{H}}-E\right)\Psi^{{JM}}(\vec{\xi}_1,\vec{\xi}_2;\vec{\mathbf{r}})=0.
\label{Diraceq}
\end{equation}\\
can be expressed as a column of  four functions
    \begin{equation}
    \Psi^{{JM}}(\vec{\xi}_1,\vec{\xi}_2;\vec{\mathbf{r}})=
    \begin{pmatrix}
        \Psi_{{LL}}(\vec{\xi}_1,\vec{\xi}_2;\vec{\mathbf{r}})\\
        \Psi_{{LS}}(\vec{\xi}_1,\vec{\xi}_2;\vec{\mathbf{r}})\\
        \Psi_{{SL}}(\vec{\xi}_1,\vec{\xi}_2;\vec{\mathbf{r}})\\
        \Psi_{{SS}}(\vec{\xi}_1,\vec{\xi}_2;\vec{\mathbf{r}})
    \end{pmatrix},
\label{column}
\end{equation}
where each of the large-large (LL), large-small (LS), small-large
(SL) and small-small (SS) components are four-component spinors and
expressed as linear combination of the harmonic-oscillator basic
functions with corresponding spin-angular parts. For the fixed total
momentum $J$, possible spin-orbital couplings include terms with
$\ell=J-1$, $\ell=J$, and $\ell=J+1$, and spins $S=0$ and $S=1$.
From Eq.(\ref{important}) and the form of the Hamiltonian of
Eq.(\ref{Hamiltonian}) one can find that the main LL- and the last
SS- components of the eigen-state contain contributions from the
same spin-orbital $(\ell,S)$ channels and differ from the LS- and
SL- components containing contributions from channels
$(\ell^{\pm},S')$, whose orbital components differ by one. In
particular, in the ground state of the para-Ps with $J^{\pi}=0^+$
the LL- and SS- components consist of only $(\ell,S)=(0,0)$
spin-orbital channel, while the LS- and SL- components contain
contributions only from the $(\ell^{\pm},S')=(1,1)$ channel. In the
case of the ortho-Ps with $J^{\pi}=1^+$, the main LL- and smallest
SS-components contain contributions from the $(\ell,S)=(0,1)$ and
$(2,1)$ spin-orbital channels, while the LS- and SL-components
contain contributions from the $(\ell^{\pm},S')=(1,0)$ and $(1,1)$
channels.

For the solution of the Dirac equation Eq.(\ref{Diraceq}) a column
of probe wave functions of Eq.(\ref{column})  is expanded in a
complete set of orthonormal harmonic oscillator states with which we
calculate the matrix elements of the Hamiltonian of
Eq.(\ref{Hamiltonian}).

 The harmonic oscillator basis functions
 $\Psi^{{JM}}_{n\ell S}(\vec{\xi}_1,\vec{\xi}_2;\vec{\mathbf{r}})$
with the variational oscillator parameter $a_0$:
\begin{equation}
    \Psi^{{JM}}_{n\ell{S}}(\vec{\xi}_1,\vec{\xi}_2;\vec{\mathbf{r}})=R_{n\ell}(a_0;r)\cdot\mathcal{Y}^{{JM}}_{\ell{S}}(\vec{\xi}_1,\vec{\xi}_2;\hat{\vec{{\mathbf{r}}}}),
\label{basis_r}
\end{equation}
where
$\mathcal{Y}^{{JM}}_{\ell{S}}(\vec{\xi}_1,\vec{\xi}_2;\hat{\vec{{\mathbf{r}}}})$
is the spherical tensor
$\mathcal{Y}^{{JM}}_{\ell{S}}(\vec{\xi}_1,\vec{\xi}_2;\hat{\vec{{\mathbf{r}}}})=\left\{\left\{\mathbf{\chi}^{{S}_1}(\vec{\xi}_1)\times\chi^{{S}_2}(\vec{\xi}_2)\right\}_{{S}}\times
\mathbf{Y}_{\ell}(\hat{\vec{\mathbf{r}}})\right\}^{{J}}_{{{M}}}$,
with ${S}_1={S}_2=\frac{1}{2}$. The radial part of the basis
function is presented as
    \begin{equation}
    R_{nl}(a_0;r)=\left[\dfrac{2(n!)}{a_0^3 \Gamma(n+\ell+\frac{3}{2})}\right]^{\frac{1}{2}}\left(\dfrac{r}{a_0}\right)^{\ell}\exp\left(-\dfrac{r^2}{2a_0^2}\right)L^{\ell+\frac{1}{2}}_n\left(r^2/a_0^2\right).
\end{equation}
where $L^{\ell+\frac{1}{2}}_n(r^2)$ are the Laguerre polynomials:
\begin{equation*}
    L^{\ell+\frac{1}{2}}_n(r^2)=\sum_{k=0}^{n}\dfrac{(-1)^k}{k!}\dfrac{\Gamma(n+\ell+3/2)}{(n-k)!\Gamma(k+\ell+3/2)}\cdot
    r^{2k}.
\end{equation*}

Accordingly, the harmonic oscillator basis functions
$\Psi^{{JM}}_{n\ell S}(\vec{\xi}_1,\vec{\xi}_2;\vec{\mathbf{r}})$
are normalized to unity. The Fourier transformation of the
coordinate-space basis functions is the same state multiplied by a
phase \cite{gutsche,gutschedis}:

\begin{equation}
    \Psi^{{JM}}_{n\ell{S}}(\vec{\mathbf{p}})=(2\pi)^{-3/2}\int d^3r\Psi^{{JM}}_{n\ell{S}}(\vec{\mathbf{r}})\exp(-i\vec{\mathbf{p}}\vec{\mathbf{r}})=
    (-i)^{2n+\ell}R_{n\ell}(a_0;p)\cdot\mathcal{Y}^{{JM}}_{\ell{S}}(\vec{\xi}_1,\vec{\xi}_2;\hat{\vec{\mathbf{p}}}),
\end{equation}
where the harmonic oscillator wave functions in momentum
representation is defined by the equation
    \begin{equation}
R_{nl}(a_0;p)=\left[\dfrac{2(n!)a_0^3}{
\Gamma(n+\ell+\frac{3}{2})}\right]^{\frac{1}{2}}(p
a_0)^\ell\exp\left(-\dfrac{p^2a_0^2}{2}\right)L^{\ell+\frac{1}{2}}_n(p^2a_0^2).
\end{equation}\\

The matrix elements of the Hamiltonian Eq.(\ref{Hamiltonian}) are
calculated in the momentum state representation
\cite{gutsche,gutschedis}, because in this case the overlaps are all
real, in contrast to the coordinate space representation, where the
contribution of the operators
$-\mathbf{I}_{2}^{(1)}\times\vec{\boldsymbol{\sigma}}_2\cdot
\hat{\vec{\mathbf{p}}}$ and
$\vec{\boldsymbol{\sigma}}_1\times\mathbf{I}_{2}^{(2)}\cdot\hat{\vec{\mathbf{p}}}$
are purely imaginary. Therefore, in momentum space, the
four-component spinors  $\Psi_{LL}$, $\Psi_{LS}$, $\Psi_{SL}$,
$\Psi_{SS}$ of Eq.(\ref{column}) are expanded in a complete set of
harmonic oscillators with corresponding coefficients according to

\begin{equation*}
 \Psi_{LL}(\vec{\xi}_1,\vec{\xi}_2;\vec{\mathbf{p}})=\sum_{n\ell S}
 a_{n\ell S}\ket{n\ell{S}{JM}}_{\mathbf{p}-space}
\end{equation*}
\begin{equation*}
    \Psi_{LS}(\vec{\xi}_1,\vec{\xi}_2;\vec{\mathbf{p}})=\sum_{n\ell S}
    b_{n\ell S}\ket{n\ell{S}{JM}}_{\mathbf{p}-space}
\end{equation*}
\begin{equation*}
\Psi_{SL}(\vec{\xi}_1,\vec{\xi}_2;\vec{\mathbf{p}})=\sum_{n\ell S}
c_{n\ell S}\ket{n\ell{S}{JM}}_{\mathbf{p}-space}
\end{equation*}
\begin{equation*}
\Psi_{SS}(\vec{\xi}_1,\vec{\xi}_2;\vec{\mathbf{p}})=\sum_{n\ell S}
d_{n\ell S}\ket{n\ell{S}{JM}}_{\mathbf{p}-space}
\end{equation*}
with \quad
$\ket{n\ell{S}{JM}}_{\mathbf{p}-space}=R_{n\ell}(a_0;p)\left\{\left\{\chi^{{S}_1}(\vec{\xi}_1)\times\chi^{{S}_2}(\vec{\xi}_2)
\right\}_{{S}}\times \mathbf{Y}_{\ell}(\hat{\vec{\mathbf{p}}})\right\}_{JM}$.\\

 A minimization of the energy is performed with the oscillator parameter  $a_0$. After
the inverse Fourier transformation, we again come to the solution of
the Dirac equation in the coordinate space:

\begin{equation*}
    \Psi_{LL}(\vec{\xi}_1,\vec{\xi}_2;\vec{\mathbf{r}})=\sum_{n\ell S}
    i^{2n+\ell}a_{n\ell{S}}R_{n\ell}(a_0;r)\mathcal{Y}^{{JM}}_{\ell{S}}(\vec{\xi}_1,\vec{\xi}_2;\hat{\vec{\mathbf{r}}})
\end{equation*}
\begin{equation*}
    \Psi_{LS}(\vec{\xi}_1,\vec{\xi}_2;\vec{\mathbf{r}})=\sum_{n\ell S}
    i^{2n+\ell}b_{n\ell{S}}R_{n \ell}(a_0;r)\mathcal{Y}^{JM}_{\ell{S}}(\vec{\xi}_1,\vec{\xi}_2;\hat{\vec{\mathbf{r}}})
\end{equation*}
\begin{equation*}
\Psi_{SL}(\vec{\xi}_1,\vec{\xi}_2;\vec{\mathbf{r}})=\sum_{n \ell S}
i^{2n+\ell}c_{n\ell{S}}R_{n\ell}(a_0;r)\mathcal{Y}^{{JM}}_{\ell{S}}(\vec{\xi}_1,\vec{\xi}_2;\hat{\vec{\mathbf{r}}})
\end{equation*}
\begin{equation*}
\Psi_{SS}(\vec{\xi}_1,\vec{\xi}_2;\vec{\mathbf{r}})=\sum_{n \ell S}
i^{2n+\ell}d_{n\ell S}
R_{n\ell}(a_0;r)\mathcal{Y}^{{JM}}_{\ell{S}}(\vec{\xi}_1,\vec{\xi}_2;\hat{\vec{\mathbf{r}}})
\end{equation*}

\section{Numerical results}

Numerical calculations have been performed with the value of the
fine-structure constant $\alpha=1/137$ for comparison with the
high-precision results of the finite-element method
\cite{scott1992}. From the form of the Dirac Hamiltonian in
Eq.(\ref{Hamiltonian}) one can find that the LS- and SL- components
of the solution of the two-body Dirac equation are identical both
for the para-Ps and ortho-Ps. For the ground and excited states of
the para-positronium with $J=0$ the main large-large (LL) and
small-small (SS) components consist of only $(\ell,S)=(0,0)$
spin-orbital channel, while the large-small (LS) and small-large
(SS) components consist of only $(\ell,S)=(1,1)$ channel. In the
case of the ortho-Ps, the large-large (LL) and small-small (SS)
components contain contributions from the $(\ell,S)=(0,1)$ and (2,1)
spin-orbital channels, while the large-small (LS) and small-large
(SL) components contain contributions from $(\ell,S)=(1,0)$ and
(1,1) channels.

\begin{figure}[htbp]
\includegraphics[width=12.8cm]{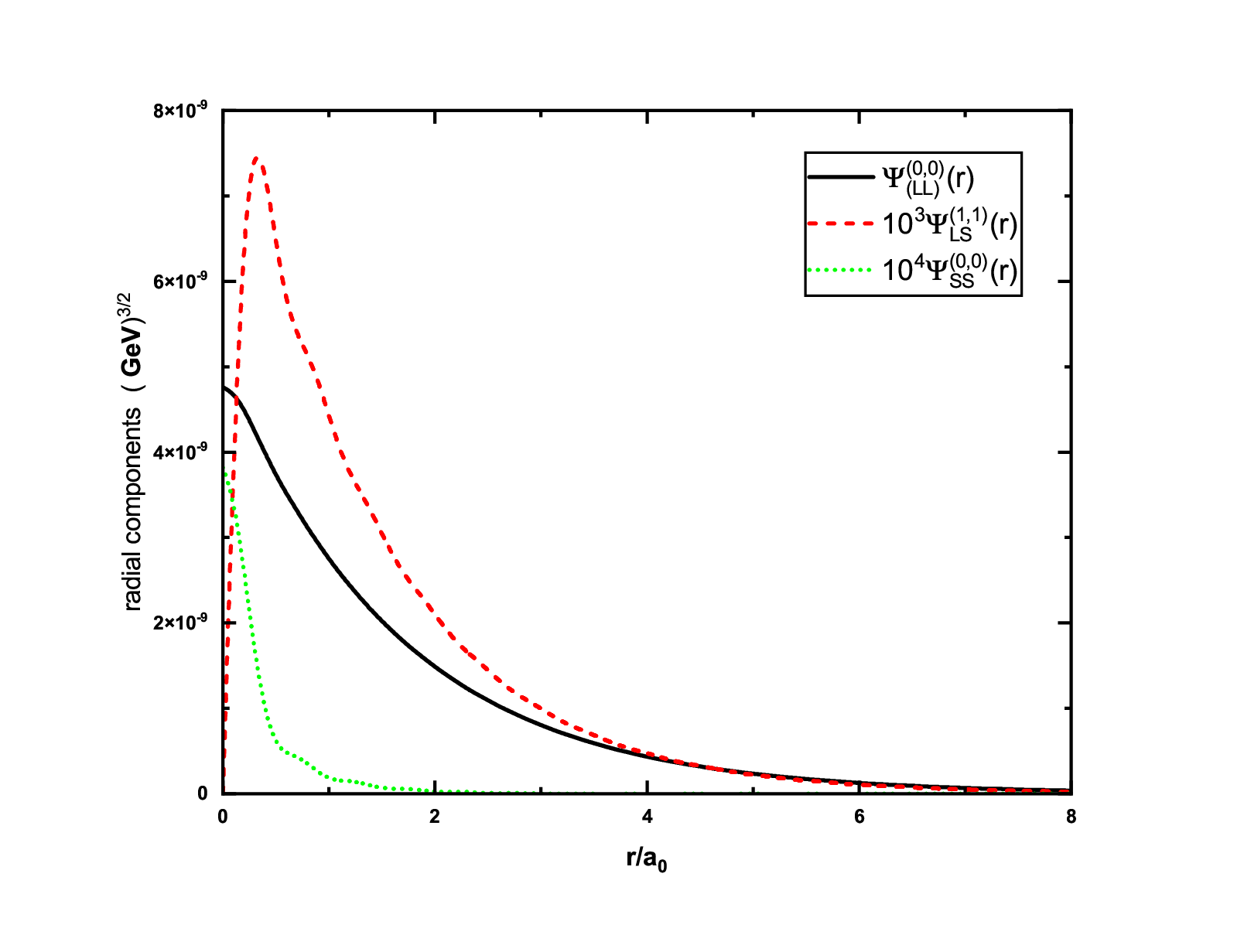}
\caption{Radial wave-functions of the large-large (LL), large-small
(LS) and small-small (SS) components of the para-Ps ground state.}
\label{fig1}
\end{figure}

\begin{figure}[htbp]
\includegraphics[width=12.8cm]{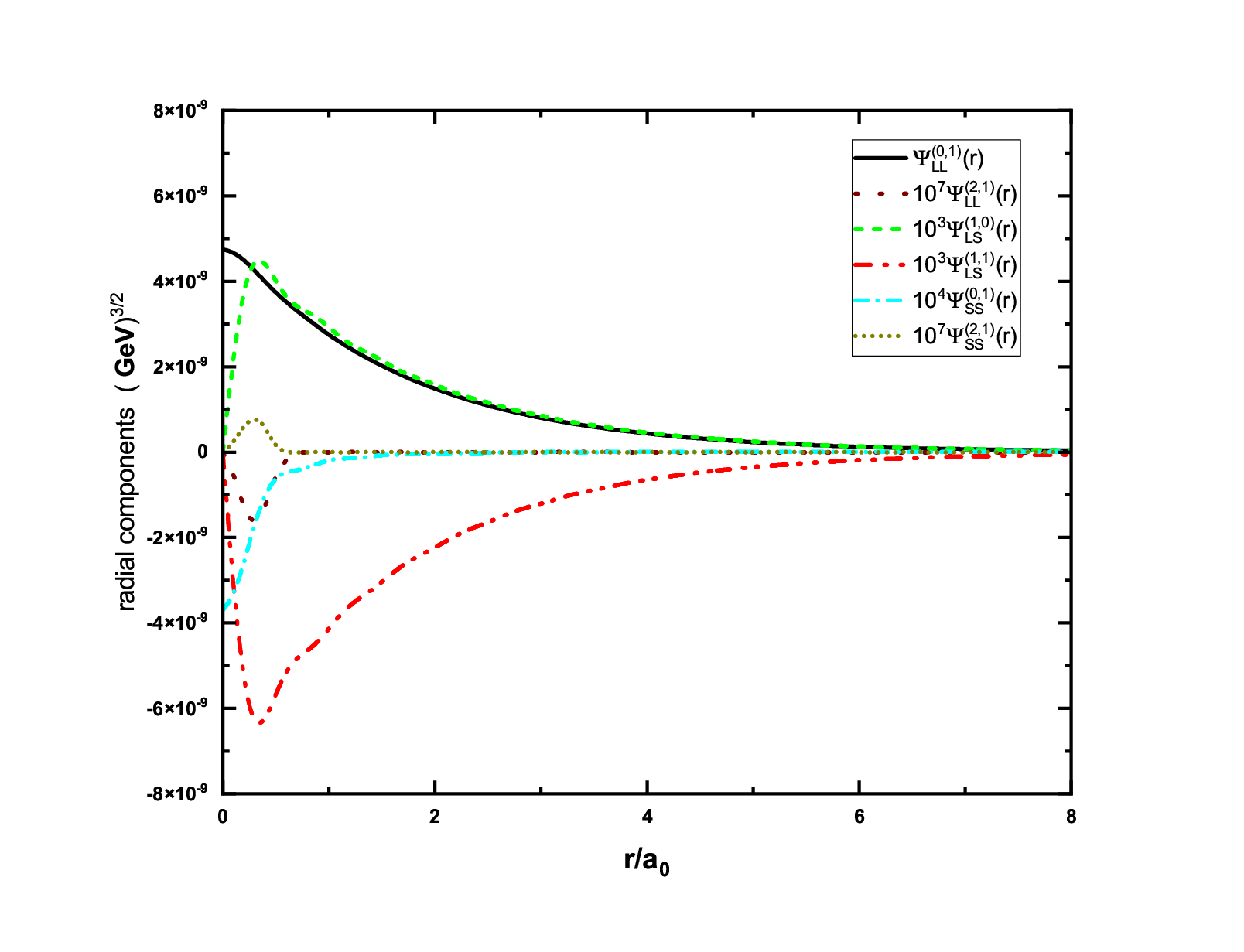}
\caption{Radial wave-functions of the large-large (LL), large-small
(LS) and small-small (SS) components of the ortho-Ps ground state.}
\label{fig2}
\end{figure}

In Fig.\ref{fig1} and Fig.\ref{fig2} the radial wave-functions of
the large-large (LL), large-small (LS) and small-small (SS)
components of the para-Ps and ortho-Ps ground states are presented,
respectively. In the upper index of the radial function notations
the corresponding spin-orbital channels are indicated as
$\Psi_{LL}^{(\ell,S)}$, $\Psi_{LS}^{(\ell,S)}$, and
$\Psi_{SS}^{(\ell,S)}$. As can be noted, the behaviors of the radial
wave functions of all the components of the para-Ps are close to the
main parts of the corresponding components of the ortho-Ps. In
particular, the corresponding LL-components are almost identical.
The values of the SS-component of the para-Ps are very close to the
absolute values of the main SS-component $\Psi_{SS}^{(0,1)}(r)$ of
the ortho-Ps. The LS-components of the ortho-Ps corresponding to
different spin-orbital channels $\Psi_{LS}^{(1,0)}(r)$ and
$\Psi_{LS}^{(1,1)}(r)$ are of the same form, but have opposite signs
and differ by about 40\% in the absolute values.

Another important result is that all the large-large, large-small ,
small-large and small-small components of the ground state wave
functions for the para-Ps and ortho-Ps do not have any singularity
around zero. This results is in contrast with the results of the
method, employed in Refs.\cite{scott1992}, where the ground state
solution of the two-body Dirac equation contains a singularity at
the origin due to the mixing with the anomalous bound state. This
problem was analyzed recently in details in Refs.
\cite{pat19,pat23}.

In Table \ref{tab1} a convergence of the numerical results of
calculations for the energy $E/2m_e$, binding energy $E_b$, the sum
of the weights of the LS- and SL- components $W=||\Psi_{(L,S)}||^2+
||\Psi_{(S,L)}||^2 $ in the para- and ortho-positronium
ground-states are presented in respect to the maximal oscillator
quantum number of basis states $n_{max}$. The optimal value of the
oscillator length $a_0=3.3\times10^5$ GeV$^{-1}$. As can be seen
from the table, all the parameters are well converged with
increasing $n_{max}$. As noted above, the further increase of
$n_{max}$ results in the collapse of the variational basis. From the
table one can see that the final result for the weights of the sum
of the large-small and small-large components are identical for the
para-Ps and ortho-Ps and estimated to be 6.656E-6. The weight of the
small-small component is of order 10$^{-12}$ in the both para-Ps and
ortho-Ps ground states.

\begin{table}[htbp]
\caption{Convergence of the numerical results for $E/2m_e$, $E_b$,
the sum of weights of the LS- and SL- components in the para- and
ortho-positronium ground-states in respect to $n_{max}$.}
{\begin{tabular}{@{}c|c|c|c|c@{}} \toprule

State & $n_{max}$ & $E/2m_e$ & $E_b$ (eV) & $W$
\\\colrule
 $^1S_0$ & 10 &0.9999935056580695  & 6.637203529 &7.467E-6  \\
 ($p$-Ps) & 20 &0.9999933597022965  & 6.786370016 &6.760E-6  \\
      & 30 & 0.9999933453955095 & 6.800991522 &6.667E-6  \\
      & 35 &0.9999933438924752  & 6.802527620 &6.659E-6  \\
      & 37 &0.9999933433568433  & 6.803075034 &6.656E-6  \\ \hline

 $^3S_1$ & 10 & 0.9999935056580694 & 6.637203529  &7.467E-6 \\
 ($o$-Ps) & 20 &0.9999933597022961  & 6.786370017  &6.756E-6  \\
      & 30 &0.9999933453955091  & 6.800991522  &6.666E-6     \\
      & 35 &0.9999933438924745  & 6.802527620  &6.660E-6   \\
      & 37 &0.9999933433568923  & 6.803074984  &6.656E-6 \\  \botrule
\end{tabular}\label{tab1}}
\end{table}

\begin{table}[htbp]
\caption{Energies of the lowest states of the para- and ortho-Ps in
comparison with the results of the finite-element method
\cite{scott1992}.} {\begin{tabular}{@{}c|c|c@{}} \toprule

     $(\ell,S,J)$ & $E/2m_e$ & $E/2m_e$ \cite{scott1992}       \\ \colrule
    (0,0,0) & 0.999 993 343 356 843 &0.999 993 340 148 538 880      \\
    (0,1,1) &0.999 993 343 356 892  &0.999 993 340 148 552 498  \\
    (1,1,0) &0.999 998 342 584 549  &0.999 998 335 009 885 854  \\
    (1,0,1) &0.999 998 342 513 543  &0.999 998 335 017 278 391  \\   \botrule
\end{tabular}\label{tab2}}
\end{table}

In Table \ref{tab2} the energies of the lowest states of the para-
and ortho-Ps are compared with the results of the finite-element
method of high accuracy \cite{scott1992}. Of course, our variational
method on the harmonic-oscillator basis is not very accurate as the
finite-element method due-to the collapse of the variational basis
at some value of $n_{max}$. The binding energies of the $p$-Ps and
$o$-Ps ground states in our calculations are estimated to be
6.803075 eV, which should be compared with the values of 6.806403 eV
for the para-Ps and 6.806354 eV for the ortho-Ps of
Ref.\cite{scott1992}. This means that the precision of the energy
calculations is of order $0.003$ eV, which is good enough for the
application of corresponding wave functions to the study of
different processes with para- and ortho-positronium.

\section{Summary}
A new theoretical method was developed to solve the two-body Dirac
equation for para- and ortho-positronium bound states. Only Coulomb
potential was included into the Dirac Hamiltonian. It is shown that
the two-body Dirac Hamiltonian can be written in the Hermitian form
and solved in the momentum state representation. Numerical results
for the energy spectrum of the para- and ortho-positronium lowest
bound states performed within the variational method using the
harmonic oscillator basis functions are in good agreement with a
high-precision finite-element method. It was shown that the para-
and ortho-Ps ground state wave functions after the Fourier
transformation to the coordinate space representation do not contain
any singularity at the origin in contrast to the traditional
bound-state solution in the literature. The weights of the sum of
the large-small and small-large components were estimated to be of
order 10$^{-6}$, while the weight of the small-small components are
of order 10$^{-12}$ for the both para- and ortho-positronium ground
states.

\section*{Acknowledgements}
The authors thank D. Baye for valuable discussions and comparison
with the results of the Lagrange-mesh method, E. Czerwinski, T.C.
Scott and A.M. Rakhimov for useful discussions of the presented
results.

\end{document}